\documentclass[thmsa,12pt]{article}
\usepackage{amsfonts}
\usepackage{amsmath,amssymb} 
\usepackage{graphicx} 
\usepackage[dvipsnames,x11names]{xcolor} 
\usepackage{hyperref} 
\hypersetup{bookmarksopen=true, bookmarksnumbered=true,
  pdfstartview={FitH}, pdfborder={0 0 0}, colorlinks=true,
  linkcolor=red, linktocpage=true, citecolor=blue, urlcolor=cyan,
  unicode=true, pdftitle={Counterterms}, pdfauthor={DJ},
  pdfsubject={Comparison of Counterterms},
  pdfkeywords={AdS, counterterms, Kounterterms}}
\usepackage{geometry} 
\usepackage{fancyhdr} 
\usepackage{parskip} 
\usepackage[nottoc,notlof,notlot]{tocbibind} 
\usepackage[titles]{tocloft} 

\definecolor{abst}{rgb}{0.6,0.366,0.2}
\definecolor{sect}{rgb}{0.2,0.3,0.7}
\definecolor{ssect}{rgb}{0.5,0.5,1.0}
\definecolor{sssect}{rgb}{0.3,0.3,0.3}
\definecolor{appsect}{rgb}{0.3,0.6,0.6}
\definecolor{eqn}{rgb}{0.2,0.3,1.0}
\definecolor{ref}{rgb}{0.0,0.0,1.0}

\numberwithin{equation}{section} 

\newcommand\arXivid[1] {\href{http://arxiv.org/abs/#1}{\tt arXiv:#1}}
\newcommand\atmp[3] {{\it Adv.\ Theor.\ Math.\
    Phys.\ }\href{http://inspirehep.net/search?ln=en&ln=en&p=find+j+''Adv.Theor.Math.Phys.,#1,#3''&of=hb&action_search=Search&sf=&so=d&rm=&rg=25&sc=0}{{\bf #1} (#2) #3}}
\newcommand\cmp[3] {{\it Commun.\ Math.\ Phys.\
  }\href{http://inspirehep.net/search?ln=en&ln=en&p=find+j+''Commun.Math.Phys.,#1,#3''&of=hb&action_search=Search&sf=&so=d&rm=&rg=25&sc=0}{{\bf  #1} (#2) #3}} 
\newcommand\cqg[3] {{\it Class.\ Quant.\ Grav.\ }\href{http://inspirehep.net/search?ln=en&ln=en&p=find+j+''Class.Quant.Grav.,#1,#3''&of=hb&action_search=Search&sf=&so=d&rm=&rg=25&sc=0}{{\bf #1} (#2) #3}}
\newcommand\jhep[3]{{\it JHEP\ }\href{http://inspirehep.net/search?ln=en&ln=en&p=find+j+''JHEP,#1,#3''&of=hb&action_search=Search&sf=&so=d&rm=&rg=25&sc=0}{{\bf #1} (#2) #3}}
\newcommand\npb[4] {{\it Nucl.\ Phys.\ }\href{http://inspirehep.net/search?ln=en&ln=en&p=j+nucl.phys.+#1#2,+#4&of=hb&action_search=Search&sf=&so=d&rm=&rg=25&sc=0}{{\bf #1 #2} (#3) #4}}
\newcommand\pr[4] {{\it Phys.\ Rev.\ }\href{http://dx.doi.org/10.1103/PhysRev#1.#2.#4}{{\bf #1 #2} (#3) #4}} 
\newcommand\pl[4] {{\it Phys.\ Lett.\
  }\href{http://inspirehep.net/search?ln=en&ln=en&p=find+j+\%22Phys.+Lett.,#1#2,#4\%22&of=hb&action_search=Search&sf=&so=d&rm=&rg=25&sc=0}{{\bf
      #1 #2} (#3) #4}}
\newcommand\phyrep[3] {{\it Phys.\
    Rept.\ }\href{http://inspirehep.net/search?ln=en&ln=en&p=find+j+\%22Phys.+Rep.,#1,#3\%22&of=hb&action_search=Search&sf=&so=d&rm=&rg=25&sc=0}{{\bf
      #1}, (#2), #3}}
\newcommand\prl[3] {{\it Phys.\ Rev.\ Lett.\
  }\href{http://inspirehep.net/search?ln=en&ln=en&p=find+j+''Phys.Rev.Lett.,#1,#3''&of=hb&action_search=Search&sf=&so=d&rm=&rg=25&sc=0}{{\bf #1} (#2) #3}}
\newcommand\prsl[4] {{\it Proc.\ Roy.\ Soc.\ Lond.\
  }\href{http://inspirehep.net/search?ln=en&ln=en&p=find+j+''Proc.Roy.Soc.Lond.,#1#2,#4''&of=hb&action_search=Search&sf=&so=d&rm=&rg=25&sc=0}{{\bf #1 #2} (#3) #4}}

\begin{document}
\title{\Huge\textbf{Conformal Mass in AdS gravity}}
\author{\textbf{Dileep
    P. Jatkar}$^1$\footnote{\href{mailto:dileep@hri.res.in}{dileep@hri.res.in}},
  \textbf{ Georgios
    Kofinas}$^2$\footnote{\href{mailto:gkofinas@aegean.gr}{gkofinas@aegean.gr}},
  \textbf{ Olivera Miskovic}$^3$\footnote{\href{mailto:olivera.miskovic@ucv.cl}{olivera.miskovic@ucv.cl}},
  \textbf{ Rodrigo
    Olea}$^4$\footnote{\href{mailto:rodrigo.olea@unab.cl}{rodrigo.olea@unab.cl}}
  \bigskip\\
  \small{ $^1$Harish-Chandra Research Institute,} \\
  \small{Chhatnag Road, Jhunsi, Allahabad 211019, India}\\
  \small{$^2$Research Group of Geometry, Dynamical Systems and Cosmology}\\
  \small{Department of Information and Communication Systems Engineering}\\
  \small{University of the Aegean, Karlovassi 83200, Samos, Greece}\\
  \small{$^3$Instituto de F\' isica, Pontificia Universidad Cat\' olica de
    Valpara\' iso,}\\
  \small{Casilla 4059, Valpara\' iso, Chile}\\
  \small{ $^4$Departamento de Ciencias F\' isicas, }\\
  \small{Universidad Andres Bello, Rep\' ublica 220, Santiago, Chile}}

\maketitle
\thispagestyle{fancy}
\vspace{3mm}
\begin{abstract}
  We show that the Ashtekar-Magnon-Das (AMD) mass and other conserved quantities are equivalent
  to the Kounterterm charges in the asymptotically AdS spacetimes that
  satisfy the Einstein equations, if we assume the same
    asymptotic fall-off behavior of the Weyl tensor as
    considered by AMD.  This therefore implies that, in all
    dimensions, the conformal mass can be directly derived from the
    bulk action and the boundary terms, which are written
    in terms of the extrinsic curvature.
\end{abstract}
\newpage
\setcounter{page}{1}
\pagestyle{plain}
\section{ Introduction}

The AdS/CFT correspondence \cite{maldacena} is an intriguing relation
between the theory of gravity defined on the anti-de Sitter (AdS) space and
the conformal field theory (CFT) defined on the boundary.  This
correspondence gives a concrete dictionary for the holographic
description of the bulk gravity in terms of the boundary data of the
conformal field theory \cite{gkp,witten,magoo}.  This, to start with,
involves matching the symmetries of the theories involved in the
correspondence.  In a diffeomorphism invariant theory, conserved
quantities are written in terms of integrals over a surface
of codimension 2, which is just a space-like slice of the space-time boundary.
These asymptotic symmetries are identified with the Killing vector fields of
the boundary space-time.

There are various definitions of these conserved quantities in the
literature.  For example, Ashtekar and Magnon showed that any
conserved charge in asymptotically AdS (AAdS) gravity in four
dimensions can be expressed as a surface integral of a quantity
involving only the electric part of the asymptotic Weyl tensor and the
asymptotic Killing field \cite{Ashtekar-Magnon}. The result was
generalized by Ashtekar and Das to higher dimensions
\cite{Ashtekar-Das}. This Ashtekar-Magnon-Das (AMD) formula for
conserved charges in AAdS spacetimes was formally proved in
Ref.\cite{Hollands-Ishibashi-Marolf} using the covariant phase space
formalism of Wald \cite{Wald-Zoupas}.  In addition, there are
approaches based on a linearized expansion around AdS background \cite{Deser-Abbott},
or on the Hamiltonian formalism (e.g.,
Henneaux-Teitelboim \cite{Henneaux-Teitelboim}), or the holographic
counterterm subtraction method of Henningson-Skenderis
\cite{Henningson-Skenderis}, developed also by Balasubramanian and
Kraus in Ref.\cite{Balasubramanian-Kraus}.  For a good comparison of
various techniques we refer the reader to
Ref.\cite{Hollands-Ishibashi-Marolf}.

The AMD method is based on the Penrose's conformal completion
techniques \cite{Penrose} that bring the boundary at infinity of the
$D$-dimensional AAdS space $\mathcal{M}$ to a finite distance. If
the physical spacetime $\mathcal{M}$ is endowed with the metric
$g_{\mu\nu}$ which obeys  the Einstein equations, then its conformal
completion consists of attaching the boundary, with topology
$\mathbb{R}\times S^{D-2}$, by means of the conformal mapping
\begin{equation}
\label{mapping}
\tilde{g}_{\mu\nu}=\Omega^2 g_{\mu\nu}.
\end{equation}
The \textit{unphysical} spacetime
$(\mathcal{\tilde{M}},\tilde{g}_{\mu\nu})$ obtained in this manner
has the\textit{\ smooth} boundary $\partial\mathcal{\tilde{M}}$ set
at a finite distance if the conformal factor $\Omega$ vanishes on
$\partial\mathcal{\tilde {M}}$ and its derivative is finite,
$\tilde{\nabla}_{\mu}\Omega=\tilde{n}_{\mu}\neq0$. The previous set
of conditions provides a definition of AAdS spacetime that will be
useful to analyze the fall-off of physical quantities in the next
section.

In particular, for asymptotic AdS spaces, which have time-like boundary, these conditions ensure that
$\Omega$ can play the role of the radial coordinate in the
neighborhood of $\partial\mathcal{\tilde{M}}$. The conformal factor
$\Omega$ is then related to the Schwarzschild-like coordinate $r$ in
the asymptotic region as $\Omega\sim1/r$ and $\tilde{n}_\mu$ is the
outward pointing normal to $\partial\mathcal{\tilde{M}}$.

The conserved quantities in the AMD approach are written in terms of
the electric part of the Weyl tensor.  The asymptotic behavior of the
Weyl tensor is therefore important in determining the expressions for
conserved charges. The fall-off behavior of the \textit{physical}
Weyl tensor $W_{\alpha\beta}^{\mu\nu}$ can be deduced as follows
\cite{Ashtekar-Das}. For the global AdS space, the Weyl tensor
vanishes identically, $W_{\alpha\beta}^{\mu\nu}=0$. Thus, a non-vacuum
state with total mass $M$ should satisfy $W_{\alpha\beta}^{\mu\nu}\sim
GM/r^{D-1}$ asymptotically, where the power factor $D-1$ is determined
purely by dimensional analysis. Using the fact that near the
boundary the conformal factor falls off as $\Omega \sim1/r$, one can
show that the fall-off behavior of the \textit{unphysical} Weyl tensor
is such that $\Omega^{4-D}\tilde{W}_{\mu\nu\alpha\beta}$ is smooth on
$\mathcal{\tilde{M}}$ and it vanishes on the boundary
$\partial\mathcal{\tilde{M}}$.  In fact, in
Ref.\cite{Hollands-Ishibashi-Marolf} it was proved that this fall-off
property is always satisfied in the AAdS spacetimes.

The \textit{electric} part of the unphysical Weyl tensor corresponds
to the Weyl tensor projected to the boundary
$\partial\mathcal{\tilde{M}}$ that is parameterized by the local
coordinates $\tilde{x}^i$,
\begin{equation}
\label{eq:elecWeyl}
\tilde{E}^i_j=\frac{1}{D-3}\,\Omega^{3-D}\,\tilde{W}^{i\mu}_{j\nu}\,\tilde
{n}_\mu\tilde{n}^\nu\,,
\end{equation}
and it is trace-free and symmetric by definition. For the spacetimes satisfying
the Einstein's equations, the leading order boundary expression of the
electric part of the unphysical Weyl tensor is also finite and
divergenceless \cite{Hollands-Ishibashi-Marolf}. As a consequence,
there exists a finite conserved charge $\mathcal{H}[\xi]$ for every
asymptotic symmetry $\xi^{i}$ for a given choice of boundary conditions,
\begin{equation}
\mathcal{H}[\xi]=-\frac{\ell}{8\pi G}\int\limits_{\tilde{\Sigma}_{\infty}}
d\tilde{\Sigma}\,\tilde{E}_{i}^{j}\,\xi^{i}\tilde{u}_{j}\,,\label{AMD}
\end{equation}
where the integral is on the spatial section $\tilde{\Sigma}_{\infty}$ of the boundary and $\ell$ is the AdS radius.
Here, $ d\tilde{\Sigma}=\sqrt{\tilde{\sigma}}\,d^{D-2}\tilde{x}$ is the surface element of
$\tilde{\Sigma}_{\infty}$ and $\tilde{u}_{j} $ is the unit timelike
normal to $\tilde{\Sigma}_\infty$.
For AAdS spacetimes, unphysical quantities correspond to regular ones at spatial infinity. In order to
write down $\mathcal{H}[\xi]$ in terms of tensors defined with the physical metric $g_{\mu \nu}$, we need to rescale
them as
\begin{align}
\tilde{n}_\mu  & =\Omega\,n_\mu\,,\nonumber\\
\tilde{u}_\mu  &=\Omega\,u_\mu\,,\nonumber\\
d\tilde{\Sigma}  & =\Omega^{D-2}d\Sigma\,,\nonumber\\
\tilde{W}_{\,\alpha\mu}^{\beta\nu}  & =\Omega^{-2}\,W_{\,\alpha\mu}^{\beta\nu}\,.
\end{align}
The Killing vector is invariant under rescaling, as it is a part of the asymptotic conformal isometry, and
hence, depends only on the conformal class of the boundary metric and not on any specific representative. This
gives rise to
\begin{equation}
\tilde{E}_i^j=\Omega^{1-D}E_i^j\,.
\end{equation}
These scaling properties are identical to those derived using
dimensional analysis since
$ \tilde{\Sigma}_\infty$ is a $(D-2)$-dimensional surface and $\tilde
{W}_{\,\beta\nu\mu}^{\alpha}=W_{\,\beta\nu\mu}^{\alpha}$. Thus, it is easy
to see that
$\tilde{E}_i^j d\tilde{\Sigma} \,\tilde{u}_j = E_i^j d\Sigma \,u_j$ and the
conformal mass (and other conserved charges) are expressed as
\begin{equation}
\mathcal{H}[\xi]=-\frac{\,\ell}{8\pi G}\,
\int\limits_{\Sigma_\infty}d\Sigma \,E_i^j\,\xi^i u_j\,.
\label{AMDfin}
\end{equation}

Since the Weyl tensor vanishes identically for the global AdS space,
the charge in the AMD approach vanishes as well.  This, in turn, implies that
it does not include the vacuum energy of the AdS space. On the other
hand, in the AdS/CFT approach, a vacuum energy is associated to
odd-dimensional AAdS spacetimes, and it matches the Casimir energy in
a holographically dual CFT. The expressions for conserved charges
agree in both approaches up to finite terms
\cite{Ashtekar-Das,Hollands-Ishibashi-Marolf,Papadimitriou}.
The AMD formula for a conserved charge in AAdS spaces gives \textit{finite},
that is regularized, quantity and moreover it takes a very simple
form because it is expressed in terms of the electric part of the
Weyl tensor only. 

When the AMD formula for conserved charges is compared with that
obtained using the holographic methods \cite{Balasubramanian-Kraus},
or the holographic renormalization techniques
\cite{Henningson-Skenderis}, the result agrees, up to two caveats:
(\textit{i}) The conserved charge derived using the holographic
methods always gives rise to the vacuum energy in odd dimensions,
whereas as stated above, the AMD charge is designed in such a way that
it vanishes in the pure AdS background, and hence, it does not contain
the vacuum energy piece by construction, and (\textit{ii}) The
holographic methods contain additional finite terms. These kind of
terms are scheme-dependent and do not contribute to the charge, after
all, the physical quantity should be independent of choice of
regularization.

We should emphasize here that the comparison between the AMD charge
and the holographic charge has been done only up to five
dimensions  \cite{Ashtekar-Das,Hollands-Ishibashi-Marolf,Papadimitriou},
partly because it becomes technically difficult to carry
out this comparison in higher dimensions. On the other hand, the
charge formula based on the Kounterterm method \cite{Olea-K} is known
in any dimension and for any Lovelock AdS gravity \cite{kofinas-olea}.
When the charges derived using the Kounterterm method are compared
with those obtained from the holographic methods for some classes of
Lovelock theories, it was shown \cite{Miskovic-Olea} that, up to the
finite counterterms, they are in agreement with each other including
the vacuum energy. This is a consequence of the fact that, in AAdS spacetimes,
the extrinsic curvature can be expanded in terms of intrinsic quantities of
the boundary \cite{Papadimitriou}.

In this paper we show that the Kounterterm charges without vacuum
energy are in \textit{exact } agreement with the AMD charges, provided
we consider the same asymptotic fall-off behavior for the Weyl tensor
as considered by AMD.  We hasten to point out here that while this
behavior of the Weyl tensor was essentially assumed by AMD, it was put
on firmer footing in \cite{Hollands-Ishibashi-Marolf}.
Combining our results with those of
  \cite{Hollands-Ishibashi-Marolf,Miskovic-Olea}, we see that the
  Kounterterm charges are natural candidates to carry out comparison
  with both the AMD charges and charges obtained using the holographic
  methods. Although direct comparison of AMD and holographic methods
  is cumbersome, the Kounterterm method has the advantage that it can
  be compared with either of them with much less technical
  difficulty.

\section{ Kounterterm charges in Einsten-Hilbert AdS gravity}

The Einstein-Hilbert AdS gravity in $D$ dimensions is described by the
action%
\begin{equation}
\label{eq:EHAdS}
I=\frac{1}{16\pi G}\int\limits_{\mathcal{M}}d^{D}x\,\sqrt{-g}\,\left(
R-2\Lambda\right)  +c_{D-1}\int\limits_{\partial\mathcal{M}}d^{D-1}%
x\,B_{D-1}\,,
\end{equation}
where $G$ is the gravitational constant in $D$ dimensions and
$\Lambda=-\left( D-1\right) \left( D-2\right) /2\ell^{2}$ is the
cosmological constant expressed in terms of the radius $\ell$ of AdS space.
The scalar curvature $R$ is constructed from the Riemann
curvature $R_{\ \nu \alpha\beta}^{\mu}$ of the spacetime, and
$B_{D-1}$, $c_{D-1}$ denote the boundary term and its coupling
constant respectively, as we will discuss below.

The Weyl tensor
\begin{equation}
\label{eq:Weyl}
W_{\alpha\beta}^{\mu\nu}=R_{\alpha\beta}^{\mu\nu}-\frac{1}{D-2}\,\left(
\delta_{\alpha}^{\mu}R_{\beta}^{\nu}-\delta_{\alpha}^{\nu}R_{\beta}^{\mu
}-\delta_{\beta}^{\mu}R_{\alpha}^{\nu}+\delta_{\beta}^{\nu}R_{\alpha}^{\mu
}\right)  +\frac{R}{\left(  D-1\right)  \left(  D-2\right)  }\,\delta
_{[\alpha\beta]}^{[\mu\nu]}
\end{equation}
vanishes for the global AdS space, which is a vacuum of the
theory. (For definition of antisymmetrized Kronecker delta, see Appendix.) This can be made explicit
by noting that the Einstein equation gives $R_{\mu\nu} = - (D-1)g_{\mu\nu}/\ell^2$ and
$R=-D(D-1)/\ell^2$. Substituting this in the expression for the Weyl
tensor, it can be written in terms of the AdS curvature,
\begin{equation}
W_{\mu\nu}^{\alpha\beta}=R_{\mu\nu}^{\alpha\beta}+\frac{1}{\ell^{2}}%
\,\delta_{[\mu\nu]}^{[\alpha\beta]}\,.\label{W}%
\end{equation}
This expression clearly vanishes for pure AdS space.

We choose the radial foliation of the manifold $\mathcal{M}$ in the local
coordinates $x^{\mu}=(r,x^{i})$, which defines the outward-pointing unit
normal $n_{\mu}=\left( n_{r},n_{i}\right) =(N,\vec{0})$ to the
boundary $\partial\mathcal{M}$. The boundary $\partial\mathcal{M}$ in
these local coordinates is placed at constant $r$.  The
line element on $\mathcal{M}$ can then be written in terms of the
Gaussian normal coordinates,
\begin{equation}
ds^{2}=g_{\mu\nu}\,dx^{\mu}dx^{\nu}=N^{2}\left(  r\right)
\,dr^{2}+h_{ij}(r,x)\,dx^{i}dx^{j}\,.\label{radial foliation}
\end{equation}
The geometry of the boundary $\partial\mathcal{M}$ is described by the
boundary metric $h_{ij}$ at a fixed, large value of $r$, the intrinsic curvature
$\mathcal{R}_{jkl}^{i}(h)$ and the extrinsic curvature
\begin{equation}
\label{eq:extrcurv}
K_{ij}=-\frac{1}{2}\,\pounds _{n}h_{ij}=-\frac{1}{2N}\,h_{ij}^{\prime}\,.
\end{equation}
Here, the prime denotes derivative along the normal direction,
which in the metric (\ref{radial foliation}) corresponds to the
radial derivative $\partial_r$. To determine the conserved
charge we need to carry out surface integration of the conserved current
on a codimension 2 surface in $\mathcal{M}$, which is an intersection
of a time-like slice with the boundary. In order to pick
this surface, we use a time-like foliation for the line element on
$\partial \mathcal{M}$ with the coordinates $x^i=(t,y^m)$ in the form
\begin{equation}
h_{ij}\,dx^{i}dx^{j}=-\hat{N}^{2}(t)dt^{2}+\sigma_{mn}\,(dy^{m}+\hat
{N}^{m}dt)(dy^{n}+\hat{N}^{n}dt)\,,\qquad\sqrt{-h}=\hat{N}\sqrt{\sigma
}\,.\label{ADMboundary}
\end{equation}
The constant time hypersurfaces $\Sigma_{r}$
are uniquely specified by the unit  normal
$u_{i}=(u_{t},u_{m})=(-\hat{N},\vec{0})$ to them.  The metric
$\sigma_{mn}$ describes the geometry of $\Sigma_{\infty}$, i.e. the
asymptotic boundary of spatial section at constant time.
The conserved charges $Q[\xi]$ of the theory, for a given set of
asymptotic Killing vectors $\{\xi\}$, are expressed as integrals over
$\Sigma_{\infty}$ (whose metric has been defined in Eq.(\ref{ADMboundary})),
\begin{equation}
Q[\xi]=\int\limits_{\Sigma_{\infty}}d^{D-2}y\,\sqrt{\sigma}\,u_j\,\xi^i\,
\left(q_i^j+q_{(0)i}^j\right) \,.\label{Qxigeneral}
\end{equation}
In general, the charge associated to the term $q_i^j$ will
  give rise to the mass and the angular momentum of the black hole
  solution.
In AdS gravity, because of the existence of black strings and planar black holes, the integration in Eq.(\ref{Qxigeneral}) over a different surface (given by a different normal $u_j$) gives rise to, e.g., tension of a black
string when the Killing vector is associated to translational invariance \cite{Traschen:2001pb}.
On the other hand, the $q_{(0)i}^j$ dependent part of the
  conserved charge is related to the vacuum energy of AAdS spaces,
  which exists only in odd dimensions.

\subsection{Even dimensions, $D=2n$}

We will now consider the boundary terms $B_{D-1}$ and the couplings
$c_{D-1}$.  In the even number of bulk dimensions, i.e. $D=2n$, the
boundary term has the form \cite{Olea-KerrBH}
\begin{align}
B_{2n-1}  & =2n\sqrt{-h}\int\limits_{0}^{1}dt\,\delta_{[i_{1}\cdots
i_{2n-1}]}^{[j_{1}\cdots j_{2n-1}]}\,K_{j_{1}}^{i_{1}}\left(  \frac{1}%
{2}\,\mathcal{R}_{j_{2}j_{3}}^{i_{2}i_{3}}-t^{2}K_{j_{2}}^{i_{2}}K_{j_{3}%
}^{i_{3}}\right)  \times\nonumber\\
& \qquad\cdots\times\left(  \frac{1}{2}\,\mathcal{R}_{j_{2n-2}j_{2n-1}%
}^{i_{2n-2}i_{2n-1}}-t^{2}K_{j_{2n-2}}^{i_{2n-2}}K_{j_{2n-1}}^{i_{2n-1}%
}\right)  \,,\label{B-even}%
\end{align}
where $\mathcal{R}_{kl}^{ij}$ is the Riemann tensor constructed from
the metric $h_{ij}$, and $K_{ij}$ is the extrinsic curvature defined
in (\ref{eq:extrcurv}).  The coupling constant, which is fixed by the
variational principle, is
\begin{equation}
c_{2n-1}=\frac{1}{16\pi G}\frac{(-1)^{n}\ell^{2n-2}}{n\left(  2n-2\right)
!}\,.\label{c2n-1}%
\end{equation}
The charge density tensor $q_i^j$ appearing in (\ref{Qxigeneral}) is
derived from (\ref{eq:EHAdS}) and (\ref{B-even}),
\begin{equation}
\label{eq:chrgdensity}
q_{i}^{j}=\frac{(-1)^{n}\,\ell^{2n-2}}{16\pi G\left(
2n-2\right)  !2^{n-2}}\,\delta_{[ i_{1}\cdots i_{2n-1}]}^{[jj_{2}\cdots
j_{2n-1}]}\,K_{i}^{i_{1}}\left[  R_{j_{2}j_{3}}^{i_{2}i_{3}}\cdots
R_{j_{2n-2}j_{2n-1}}^{i_{2n-2}i_{2n-1}}-\frac{(-1)^{n-1}}{\ell^{2n-2}}%
\,\delta_{[ j_{2}j_{3}]}^{[i_{2}i_{3}]}\cdots\delta_{[
j_{2n-2}j_{2n-1}]}^{[i_{2n-2}i_{2n-1}]}\right]  \,,
\end{equation}
and $q_{(0)i}^{j}=0$ for even dimensions.
The expression of the charge density tensor can be put in a convenient
form using the algebraic identity
\begin{equation}
b^{n-1}-(-a)^{n-1}=\left(  n-1\right)  \left(  b+a\right)  \int\limits_{0}%
^{1}du\,\left[  u\left(  b+a\right)  -a\right]  ^{n-2}\,.\label{b-a}%
\end{equation}
This allows us to write an integral representation for the charge
density tensor $q_{i}^{j}$ as%
\begin{align}
q_{i}^{j}  & =\frac{(n-1)( -1)^n\,\ell^{2n-2}}{16\pi G\left(
2n-2\right)  !2^{n-2}}\,\delta_{[i_{1}\cdots i_{2n-1}]}^{[jj_{2}\cdots
j_{2n-1}]}\,K_{i}^{i_{1}}\left(  R_{j_{2}j_{3}}^{i_{2}i_{3}}+\frac{1}{\ell
^{2}}\,\delta_{[j_{2}j_{3}]}^{[i_{2}i_{3}]}\right)  \times
\nonumber\\
& \times\int\limits_{0}^{1}du\left[  \left(  1-u\right)  \,R_{j_{4}j_{5}%
}^{i_{4}i_{5}}-\frac{u}{\ell^{2}}\,\delta_{[ j_{4}j_{5}]}^{[i_{4}i_{5}%
]}\right]  \cdots\left[  \left(  1-u\right)  \,R_{j_{2n-2}j_{2n-1}}%
^{i_{2n-2}i_{2n-1}}-\frac{u}{\ell^{2}}\,\delta_{[ j_{2n-2}j_{2n-1}%
]}^{[i_{2n-2}i_{2n-1}]}\right]  \,. \label{q even}
\end{align}
Let us now look at the radial dependence of the quantities appearing
in the expression of the conserved charge.  We will first look at the
Weyl tensor, significance of which will become clear momentarily.  The
fall-off of the boundary components $W_{j_{2}j_{3}}^{i_{2}i_{3}}$ of
the Weyl tensor (\ref{W}) can be deduced as follows
\cite{Ashtekar-Das}.  Recall that for the global AdS space the Weyl tensor
vanishes identically and as a result we have
$W_{j_{2}j_{3}}^{i_{2}i_{3}}=0$.  For a non-vacuum state with
total mass $M$, the Weyl tensor will be nonvanishing and in general
its asymptotic behavior will be $W_{j_{2}j_{3}}^{i_{2}i_{3}}\sim
GM/r^{s}$, where $s$ is a number that will be determined using
dimensional analysis.  In the natural units, the dimensions are
$[G]=($length$)^{D-2}$ and $[M]=($length$)^{-1}$.  Since the Weyl
tensor contains two derivatives of the metric, it is $[W]=($length$)^{-2}$.
Combining these, we find $s=D-1 $. Thus, in the AMD method, the fall-off
of the Weyl tensor is typically \cite{Hollands-Ishibashi-Marolf}
\begin{equation}
W_{j_{2}j_{3}}^{i_{2}i_{3}}=\mathcal{O}(1/r^{D-1})\,.\label{Weyl fall-off}
\end{equation}

Furthermore, using the asymptotic behavior of the metric (\ref{ADMboundary})
for AAdS spaces \`{a} la AMD, i.e. $g_{\mu\nu}\sim\Omega^2\sim1/r^2$, we can deduce
that the radial dependence of the unit normal that generates the foliation
(\ref{ADMboundary}) is
\begin{equation}
u_{t}=\mathcal{O}(r)\,,
\end{equation}
and the radial dependence of the determinant of the metric on
$\Sigma_\infty$ is
\begin{equation}
\sqrt{\sigma}=\mathcal{O}(r^{D-2})\,.
\end{equation}
These fall-off behaviors uniquely determine the behavior of the charge
tensor.  Notice that the expression of the charge tensor $q_i^j$
contains at least one power of the Weyl tensor.  The $r$ dependence of
terms in the conserved charge cancels out if only the leading terms in
the expansion of the Riemann tensor $R_{kl}^{ij}$ and the extrinsic
curvature $K_{j}^{i}$ are taken into account.  The subleading terms in
the expansion of the Riemann tensor $R_{kl}^{ij}$ and the extrinsic
curvature $K_{j}^{i}$ do not contribute to the conserved charge.  This
corresponds to the following substitutions in Eq.(\ref{q even}),
\begin{equation}
R_{kl}^{ij}=-\frac{1}{\ell^2}\,\delta_{[kl]}^{[ij]} +\mathcal{O}(1/r^2)\,,
\qquad K_{j}^{i}=-\frac{1}{\ell}\,\delta_j^i +\mathcal{O}(1/r^2)\,.
\label{R and K}
\end{equation}
This substitution in (\ref{q even}) completely eliminates its
dependence on the parameter $u$, and as a result, the
integration over $u$ in Eq.(\ref{q even}) becomes trivial,
\begin{align}
q_i^j  & =\frac{(n-1)(-1)^n\,\ell^{2n-2}}{16\pi G\left(2n-2\right)!2^{n-2}}\,
\delta_{[i_{1}\cdots i_{2n-1}]}^{[jj_{2}\cdots j_{2n-1}]}\,\left(-\frac{1}{\ell^2}\,\delta_i^{i_1}\right)\,
W_{j_{2}j_{3}}^{i_{2}i_{3}}  \times
\nonumber\\
& \times\int\limits_0^1 du \left( -\frac{1}{\ell^{2}}\right)^{n-2}  \, \delta_{[ j_{4}j_{5}]}^{[i_{4}i_{5}]}
\cdots\delta_{[ j_{2n-2}j_{2n-1}]}^{[i_{2n-2}i_{2n-1}]}+\mathcal{O}(1/r^{D+1}) \,,
\end{align}
and we obtain for the charge density tensor,
\begin{equation}
\label{eq:qweyl}
q_i^j=-\frac{(n-1)\,\ell}{16\pi G\left(  2n-2\right)  !2^{n-2}}\,
\delta_{[ ii_{2}\cdots i_{2n-1}]}^{[jj_{2}\cdots j_{2n-1}]}\,
W_{j_{2}j_{3}}^{i_{2}i_{3}}\,\delta_{[ j_{4}j_{5}]}^{[i_{4}i_{5}]}
\cdots\delta_{[ j_{2n-2}j_{2n-1}]}^{[i_{2n-2}i_{2n-1}]}
+ \mathcal{O}(1/r^{D+1})\,.
\end{equation}
Taking into account the fall-off behavior of the Weyl tensor, it
  is clear that the asymptotic radial dependence of $q_i^j$ goes
  as $1/r^{D-1}$.  This behavior of $q_i^j$ gives rise to the
  \emph{finite} conserved charge (\ref{Qxigeneral}), as expected.

Using multiplicative properties of antisymmetrized Kronecker deltas
we can contract them in (\ref{eq:qweyl}) to get a simple expression
for the charge tensor,
\begin{align}
q_{i}^{j}  & =-\frac{\,\ell}{32\pi G\left(  2n-3\right)  }\,\delta_{[
ii_{2}i_{3}]}^{[jj_{2}j_{3}]}\,W_{j_{2}j_{3}}^{i_{2}i_{3}}\nonumber\\
& =-\frac{\,\ell}{16\pi G\left(  2n-3\right)  }\,\left(  \delta_{i}^{j}%
\,W_{kl}^{kl}-2W_{ki}^{kj}\right)  \,,\label{QWeyl}
\end{align}
where we have neglected the $\mathcal{O}(1/r^{D+1})$ terms that
    fall off rapidly at the boundary and hence will give vanishing
    contribution to the charge.
We will now use the fact that the Weyl tensor is traceless, that is
$W_{\mu\beta }^{\mu\alpha}=0$, which can be used to write down
relations between the components of the Weyl tensor,
\begin{equation}
W_{kr}^{kr}=0\,,\qquad W_{rj}^{ri}+W_{kj}^{ki}=0\,.
\end{equation}
Taking trace of the second equation, we also find $W_{ki}^{ki}=0$.
As a consequence of these relations we further simplify the expression
of $q_i^j$.  It is easy to see that the first term in Eq.(\ref{QWeyl})
vanishes, and the second term can be rewritten in terms of $W_{ri}^{rj}$,
\begin{equation}
q_{i}^{j}=-\frac{\ell}{8\pi G\left(2n-3\right)  }\,W_{ri}^{rj}\,.
\end{equation}
On the other hand, the electric part of the Weyl tensor $E_{i}^{j}$,
defined in terms of the normal $n_{\mu}$ to the boundary, reads
\begin{equation}
E_{i}^{j}=\frac{1}{D-3}\,W_{i\mu}^{j\nu}\,n^{\mu}n_{\nu}\,.\label{E}
\end{equation}
The Kounterterm charge density tensor can now be written in terms the
electric part of the Weyl tensor
\begin{equation}
q_{i}^{j}=-\frac{\,\ell}{8\pi G}\,E_{i}^{j}\,.
\end{equation}
Substituting this in the conserved charge formula we get
\begin{equation}
Q[\xi] =-\frac{\,\ell}{8\pi G}\,\int\limits_{\Sigma_{\infty}}d^{D-2}%
y\,\sqrt{\sigma}\,E_{i}^{j}\,\xi^{i}\,u_{j}
 =-\frac{\,\ell}{8\pi G}\,\int\limits_{\Sigma_\infty}d\Sigma\,
 E_i^j\,\xi^i\,u_j\,,
\end{equation}
where
\begin{equation}
d\Sigma=d^{D-2}y\,\sqrt{\sigma}\,\,.
\end{equation}
Therefore, the conserved quantities $Q[\xi]$ coming from Kounterterms
regularization in $D=2n$ are the same as Eq.(\ref{AMDfin}), that is
the Ashtekar-Magnon-Das formula (\ref{AMD}).  Thus, we conclude
\begin{equation}
\mathcal{H}[\xi]=Q[\xi]\,.
\end{equation}
In four dimensions, the equivalence between the conformal mass
  and the Kounterterm charge (\ref{q even}) was established using the
  asymptotic expansion of the metric in Ref.\cite{Miskovic-Olea-Top}.

\subsection{Odd dimensions, $D=2n+1$}

The boundary term that regularizes the Einstein-Hilbert AdS gravity in odd
dimensions is given in terms of two parametric integrations as \cite{Olea-K}
\begin{eqnarray}
B_{2n} &=&2n\sqrt{-h}\int\limits_0^1 du\int\limits_0^u ds\,\delta_{[i_1
\cdots i_{2n}]}^{[j_{1}\cdots j_{2n}]}\,K_{j_1}^{i_1}\delta _{j_{2}}^{i_{2}}
\left( \frac{1}{2}\,\mathcal{R}_{j_{3}j_{4}}^{i_{3}i_{4}}-u^{2}K_{j_{3}}^{i_{3}}K_{j_{4}}^{i_{4}}
+\frac{s^{2}}{\ell _{e\!f\!f}^{2}}\,\delta _{j_{3}}^{i_{3}}\delta_{j_{4}}^{i_{4}}\right) \times   \nonumber \\
&&\cdots \times \left( \frac{1}{2}\,\mathcal{R}
_{j_{2n-1}j_{2n}}^{i_{2n-1}i_{2n}}-u^{2}K_{j_{2n-1}}^{i_{2n-1}}K_{j_{2n}}^{i_{2n}}+
\frac{u^{2}}{\ell _{e\!f\!f}^{2}}\,\delta _{j_{2n-1}}^{i_{2n-1}}\delta
_{j_{2n}}^{i_{2n}}\right) \,,
\end{eqnarray}
where the corresponding coupling constant, obtained from the
variational principle, has the form
\begin{equation}
c_{2n}=-\frac{1}{16\pi G}\frac{\ell^{2n-2}}{n(2n-1)!}\,\left[
\int\limits_{0}^{1}du\,\left(  u^{2}-1\right)  ^{n-1}\right]  ^{-1}\,.
\label{nc2n}
\end{equation}
This integral representation will be useful later.

The Noether charge is written as an integral over the sum of two tensors as
in Eq.(\ref{Qxigeneral}).  The term $q_{(0)i}^{j}$ in
Eq.(\ref{Qxigeneral}) no longer vanishes in odd dimensions. The first
part takes the form
\begin{align}
q_{i}^{j}  & =\frac{1}{16\pi G\left( 2n-1\right)  !2^{n-2}}\,
\delta_{[i_{1}i_{2}\cdots i_{2n}]}^{[jj_{2}\cdots j_{2n}]}\,K_{i}^{i_{1}}
\delta_{j_{2}}^{i_{2}} \left[ \rule{0in}{0.30in}\delta_{[ j_{3}j_{4}]}^{[i_{3}i_{4}%
]}\cdots\delta_{[ j_{2n-1}j_{2n}]}^{[i_{2n-1}i_{2n}]}\right. \nonumber\\
& +\left.  16\pi G\left(  2n-1\right)  !\,nc_{2n}\int\limits_{0}^{1}du\left(
R_{j_{3}j_{4}}^{i_{3}i_{4}}+\frac{u^{2}}{\ell^{2}}\,\delta_{[ j_{3}%
j_{4}]}^{[i_{3}i_{4}]}\right)  \cdots\left(  R_{j_{2n-1}j_{2n}}^{i_{2n-1}%
i_{2n}}+\frac{u^{2}}{\ell^{2}}\,\delta_{[ j_{2n-1}j_{2n}]}%
^{[i_{2n-1}i_{2n}]}\right)  \right] \label{q(2n-1)}\,.
\end{align}
The second part, namely $q_{(0)i}^{j}$, of the charge represents a
covariant formula for the vacuum energy for any asymptotically AdS
spacetime%
\begin{align}
q_{(0)i}^{j}  & =nc_{2n}\,\int\limits_{0}^{1}du\,u\,\delta_{[
ki_{2}\cdots i_{2n}]}^{[jj_{2}\cdots j_{2n}]}\left(  K_{i}^{k}\delta_{j_{2}%
}^{i_{2}}+K_{j_{2}}^{k}\delta_{i}^{i_{2}}\right)  \left(  \frac{1}%
{2}\,\mathcal{R}_{j_{3}j_{4}}^{i_{3}i_{4}}-u^{2}K_{j_{3}}^{i_{3}}K_{j_{4}%
}^{i_{4}}+\frac{u^{2}}{\ell^{2}}\,\delta_{j_{3}}^{i_{3}}\delta_{j_{4}}^{i_{4}%
}\right)  \times\cdots\nonumber\\
& \qquad\qquad\cdots\times\left(  \frac{1}{2}\,\mathcal{R}_{j_{2n-1}j_{2n}%
}^{i_{2n-1}i_{2n}}-u^{2}K_{j_{2n-1}}^{i_{2n-1}}K_{j_{2n}}^{i_{2n}}+\frac
{u^{2}}{\ell^{2}}\,\delta_{j_{2n-1}}^{i_{2n-1}}\delta_{j_{2n}}^{i_{2n}%
}\right)  \,.\label{qzero}%
\end{align}
The latter term $q_{(0)i}^{j}$, which corresponds to the vacuum energy
of the asymptotically AdS space, is not considered in the AMD method.
We will therefore restrict our current discussion to the charge
(\ref{q(2n-1)}) and ignore the term (\ref{qzero}).

In order to perform a comparison, we take the charge (\ref{q(2n-1)}) and
express it as
\begin{align}
q_{i}^{j}  & =\frac{nc_{2n}}{2^{n-2}}\,\int\limits_{0}^{1}du\,\delta_{[
i_{1}i_{2}\cdots i_{2n}]}^{[jj_{2}\cdots j_{2n}]}\,K_{i}^{i_{1}}\delta_{j_{2}%
}^{i_{2}}\left[  \left(  R_{j_{3}j_{4}}^{i_{3}i_{4}}+\frac{u^{2}}{\ell^{2}%
}\,\delta_{[ j_{3}j_{4}]}^{[i_{3}i_{4}]}\right)  \cdots\left(
R_{j_{2n-1}j_{2n}}^{i_{2n-1}i_{2n}}+\frac{u^{2}}{\ell^{2}}\,\delta_{[
j_{2n-1}j_{2n}]}^{[i_{2n-1}i_{2n}]}\right)  \right. \nonumber\\
&\quad\quad\quad\quad\quad\quad\quad\quad\quad\quad\quad\quad\quad\quad
 \left.  -\left(  \frac{u^{2}-1}{\ell^{2}}\,\right)  ^{n-1}\rule{0in}{0.26in}%
\delta_{[ j_{3}j_{4}]}^{[i_{3}i_{4}]}\cdots\delta_{[
j_{2n-1}j_{2n}]}^{[i_{2n-1}i_{2n}]}\right]  \,,\label{odd-dq}
\end{align}
with the explicit form (\ref{nc2n}) of the constant $c_{2n}$.

As in the even-dimensional case, we can show that the charge is
factorizable over $\left( R\!+\!\frac{1}{\ell^{2}}\delta^{[2]}\right)$,
by employing the formula (\ref{b-a}). To do this, we introduce another
integration parameter $s$ and choose the coefficients
$b=R+\frac{u^{2}}{\ell^{2}}\,\delta^{[2]}$ and
$a=\frac{1-u^{2}}{\ell^{2}}\,\delta^{[2]}$.  Using this integral
representation we can rewrite the charge density tensor (\ref{odd-dq})
as
\begin{align}
q_{i}^{j}  & =n\left(  n-1\right)  \,\frac{c_{2n}}{2^{n-2}}\,\delta_{[
i_{1}i_{2}\cdots i_{2n}]}^{[jj_{2}\cdots j_{2n}]}\,K_{i}^{i_{1}}\delta_{j_{2}%
}^{i_{2}}\,\left(  R_{j_{3}j_{4}}^{i_{3}i_{4}}+\frac{1}{\ell^{2}}%
\,\delta_{[ j_{3}j_{4}]}^{[i_{3}i_{4}]}\right)  \times\nonumber\\
& \int\limits_{0}^{1}du\int\limits_{0}^{1}ds\,\left[  s\left(  R_{j_{5}j_{6}%
}^{i_{5}i_{6}}+\frac{1}{\ell^{2}}\,\delta_{[ j_{5}j_{6}]}^{[i_{5}i_{6}%
]}\right)  +\frac{u^{2}-1}{\ell^{2}}\,\delta_{[ j_{5}j_{6}]}%
^{[i_{5}i_{6}]}\right]  \times\cdots\nonumber\\
& \cdots\times\left[  s\left(  R_{j_{2n-1}j_{2n}}^{i_{2n-1}i_{2n}}+\frac
{1}{\ell^{2}}\,\delta_{[ j_{2n-1}j_{2n}]}^{[i_{2n-1}i_{2n}]}\right)
+\frac{u^{2}-1}{\ell^{2}}\,\delta_{[ j_{2n-1}j_{2n}]}^{[i_{2n-1}i_{2n}
]}\right]  .
\label{q odd}
\end{align}
Notice that, in three dimensions, $q_{i}^{j}$ is zero, what is
a reflection of the fact that the Weyl tensor vanishes identically.
Therefore, the mass of AAdS black holes in this case comes from Eq.(\ref{qzero}).

For $n>1$, using (\ref{W}), we can rewrite (\ref{q odd}) in terms of the Weyl
tensor.  When we utilize the fall-off behavior of the Weyl
tensor given in Eq.(\ref{Weyl fall-off}), we are essentially forced to
consider only the leading-order terms in the asymptotic expansion of
the curvature tensors, that is Eq.(\ref{R and K}).  We can therefore
express the charge density tensor (\ref{odd-dq}) as
\begin{align}
q_i^j  & =n(n-1)\,\frac{c_{2n}}{2^{n-2}}\,\delta_{[i_1 i_2\cdots
  i_{2n}]}^{[jj_{2}\cdots j_{2n}]}\,
\left( -\frac{1}{\ell}\,\delta_i^{i_1}\right)\delta_{j_2}^{i_2}\,
W_{j_3 j_4}^{i_3 i_4}\times\nonumber\\
& \times\int\limits_0^1 du \int\limits_0^1 ds\,
\left(\frac{u^2-1}{\ell^2}\right)^{n-2}\,
\delta_{[j_5 j_6]}^{[i_5 i_6]} \cdots \delta_{[ j_{2n-1}
  j_{2n}]}^{[i_{2n-1}i_{2n}]}+\mathcal{O}(1/r^{D+1})\,. \label{simplified-q}
\end{align}
Since the integrand in (\ref{simplified-q}) is independent of $s$, we
can carry out the trivial integration over the parameter $s$, and the
charge is rewritten as
\begin{equation}
q_i^j  =\frac{(n-1)\ell}{16\pi
  G\,(2n-1)!\,2^{n-2}}\frac{\int\limits_{0}^{1}du\left(u^2-1\right)^{n-2}}
{\int\limits_{0}^{1}du\left(u^2-1\right)  ^{n-1}}\,
 \delta_{[ ii_{2}\cdots i_{2n-1}]}^{[jj_{2}\cdots j_{2n-1}]}
\,W_{j_{2}j_{3}}^{i_{2}i_{3}}\,\delta_{[ j_{4}j_{5}]}^{[i_{4}i_{5}
]}\cdots\delta_{[ j_{2n-2}j_{2n-1}]}^{[i_{2n-2}i_{2n-1}]}\,,
\end{equation}
where we again omit writing the $\mathcal{O}(1/r^{D+1})$ terms that
vanish on the boundary.

Using the properties of product of antisymmetrized Kronecker deltas listed
in the Appendix and the relation
\begin{equation}
\frac{\int\limits_{0}^{1}du \left( u^2-1\right)^{n-2}}
{\int\limits_{0}^{1}du\left(u^2-1\,\right)^{n-1}
}=- \frac{2n-1}{2(n-1)}\,,
\end{equation}
we can further simplify the formula to
\begin{equation}
q_{i}^{j}=- \frac{\,\ell\,}{32\pi G\left( 2n-2\right)!}\,(2n-3)!\,\delta
_{[ ii_{2}i_{3}]}^{[jj_{2}j_{3}]}\,W_{j_{2}j_{3}}^{i_{2}i_{3}}\,.\label{eq:36}
\end{equation}
The above formula (\ref{eq:36}) is equivalent to
\begin{align}
q_{i}^{j}  & =-\frac{\,\ell}{32\pi G\left(  2n-2\right)  }\,\delta_{[
ii_{2}i_{3}]}^{[jj_{2}j_{3}]}\,W_{j_{2}j_{3}}^{i_{2}i_{3}}\nonumber\\
& =\frac{\,\ell}{8\pi G\left(  2n-2\right)  }\,W_{ki}^{kj}\,.
\end{align}
This form of the charge density tensor is equivalent to that in
  even dimensions.  Hence it can be again written in terms of the
electric part of the Weyl tensor as
\begin{equation}
q_{i}^{j}=-\frac{\,\ell}{8\pi G}\,E_i^j\,.
\end{equation}

Thus, we find that in the odd number of dimensions too the
  conformal mass and the mass derived from the Kounterterms are related
in a similar way, namely,
\begin{equation}
\mathcal{H}[\xi]=Q[\xi]\,.
\end{equation}
This proves that the conserved quantities, derived from
the action supplemented with Kounterterms,
are in exact agreement with the AMD charges (\ref{AMD}) in any dimension.

\section{ Conclusions}

We have provided an explicit comparison between conformal mass
in AAdS gravity and Kounterterm charges in all dimensions.

The agreement between these different notions of conserved
quantities in AdS gravity seems to indicate that most of the
holographic information of AAdS spacetimes is encoded in the
electric part of the Weyl tensor. The only difference with respect
to holographic charges appears in the odd-dimensional case, where
there is a piece that gives rise to the vacuum energy.

AMD charges are obtained from the asymptotic resolution of the
bulk field equations to the relevant order in  the
conformal factor $\Omega$, such that the addition of boundary terms
to the action does not play any role in their derivation.
The proof given here links the definition of conformal mass to the
addition of extrinsic counterterms in all dimensions, and the
fact that the variation of the total action is factorizable over the Weyl tensor.

The Kounterterm regularization can be extended to the Einstein-Gauss-Bonnet
gravity for all values of the Gauss-Bonnet coupling, where two
AdS branches can be defined. The charge formulae for this theory can
be factorized in a similar fashion as in Eqs.(\ref{q even}) and (\ref{q odd}) for
the Einstein gravity, but the relation (\ref{W}) does not
hold in that case.  However, it can be seen that Eq.(\ref{E})
is proportional to the boundary Weyl tensor in Einstein-Hilbert gravity. Thus, one can use
the fact that, at a linearized level, charge expressions for Einstein-Gauss-Bonnet are proportional to the
boundary Weyl tensor to work out a conformal mass definition in this gravity theory.

\section*{Acknowledgments}

We thank A.~Ashtekar for clarifying comments related to conformal
  mass method.  This work was supported by the Chilean FONDECYT Grants
  No.~1131075 and No.~1110102. O.M. also thanks DII-PUCV for support
  through the project No.~123.711/2011. The work of R.O. is financed
  in part by the UNAB grant No.~DI-551-14/R. D.P.J. would
  like to thank Departamento de Ciencias F\' isicas, UNAB and
  Instituto de F\' isica, PUCV for warm hospitality.  The work of D.P.J.
  is partly supported by DAE project 12-R\& D-HRI-5.02-0303.

\appendix{}

\section{Conventions\label{Notation}}

We will list here our notation and conventions used in the main text.
The antisymmetrized Kronecker delta is defined as
\begin{equation}
  \label{eq:3}
  \delta _{\left[ \mu _{1}\cdots \mu _{m}\right] }^{\left[ \nu _{1}\cdots \nu
_{m}\right] }:=\left\vert
\begin{array}{cccc}
\delta _{\mu _{1}}^{\nu _{1}} & \delta _{\mu _{1}}^{\nu _{2}} & \cdots &
\delta _{\mu _{1}}^{\nu _{m}} \\
\delta _{\mu _{2}}^{\nu _{1}} & \delta _{\mu _{2}}^{\nu _{2}} &  & \delta
_{\mu _{2}}^{\nu _{m}} \\
\vdots &  & \ddots &  \\
\delta _{\mu _{m}}^{\nu _{1}} & \delta _{\mu _{m}}^{\nu _{2}} & \cdots &
\delta _{\mu _{m}}^{\nu _{m}}%
\end{array}%
\right\vert \,.
\end{equation}

In $d$ dimensions, a contraction of $k\leq p$ indices in the Kronecker delta
of rank $p$ produces a delta of rank $p-k$,
\begin{equation}
\delta _{\left[ \mu _{1}\cdots \mu _{k}\cdots \mu _{p}\right] }^{\left[ \nu
_{1}\cdots \nu _{k}\cdots \nu _{p}\right] }\,\delta _{\nu _{1}}^{\mu
_{1}}\cdots \delta _{\nu _{k}}^{\mu _{k}}=
\frac{\left( d-p+k\right) !}{\left( d-p\right) !}\,
\delta _{\left[ \mu _{k+1}\cdots \mu _{p}\right] }^{\left[ \nu _{k+1}\cdots \nu _{p}\right] }\,.
\end{equation}

We work with the Gaussian normal coordinates in which the metric takes
the form
\begin{equation}
  \label{eq:2}
  ds^2 = g_{\mu\nu}dx^\mu dx^\nu = N^2(r) dr^2 + h_{ij}(r,x) dx^i
  dx^j\, .
\end{equation}

The Riemann curvature tensor is defined as
\begin{equation}
R_{\;\;\nu \lambda \rho }^\mu =\partial _\lambda\Gamma _{\nu \rho}^\mu
-\partial _\rho \Gamma _{\nu \lambda }^\mu
+\Gamma _{\sigma\lambda }^\mu \Gamma _{\nu \rho }^\sigma
-\Gamma _{\sigma \rho }^\mu\Gamma _{\nu \lambda }^\sigma \,,
\end{equation}
and the extrinsic curvature of the boundary $\partial \mathcal{M}$ (with the
boundary metric $h_{ij}$) reads
\begin{equation}
K_{ij}=-\frac{1}{2}\,\pounds _{n}h_{ij}=-\frac{1}{2N}\,h_{ij}^\prime \,.
\end{equation}

The outward-pointing unit normal of $\partial \mathcal{M}$ is the
space-like unit vector, $g_{\mu\nu}\,n^\mu n^\nu =1$, with the
components $n_\mu =(n_r,n_i)=(N,\vec{0})$.

The line element on $\partial \mathcal{M}$ has the ADM form
\begin{equation}
h_{ij}\,dx^i dx^j =-\hat{N}^2(t)\,dt^2+\sigma _{mn}\,(dy^m
+\hat{N}^m dt)(dy^n +\hat{N}^n dt)\,.
\end{equation}

The unit normal of the surface $\Sigma _r$ (endowed
with the metric $\sigma _{mn}$) is the time-like unit vector,
$h_{ij}\,u^i u^j =-1$, with components
$u_i=(u_t,u_m)=(-\hat{N},\vec{0})$.

\end{document}